\documentclass[aps,twocolumn,tightenlines,nofootinbib,superscriptaddress]{revtex4-2}
\usepackage[utf8]{inputenc}
\usepackage{graphicx}
\usepackage{mwe}
\usepackage{mathrsfs}
\usepackage{amsfonts}
\usepackage{setspace}
\usepackage{cellspace}
\usepackage{amsmath,amssymb,bm}
\usepackage[colorlinks=true,linkcolor=blue]{hyperref}
\usepackage{xcolor}
\usepackage{epsfig}
\usepackage{slashed}
\usepackage{hhline,multirow,tabularx}  %
\usepackage{dcolumn}    %
\usepackage{url}        %
\usepackage{braket}     %
\usepackage{makecell}
\usepackage{bbm}

\DeclareMathAlphabet{\mathdutchcal}{U}{dutchcal}{m}{n}
\SetMathAlphabet{\mathdutchcal}{bold}{U}{dutchcal}{b}{n}
\DeclareMathAlphabet{\mathdutchbcal}{U}{dutchcal}{b}{n}

\begin{document}

\title{Uncover the correlation between jet energy correlators and multiplicity fluctuations}

\author{Pi Duan}
\email{duanpi@mails.ccnu.edu.cn}
\affiliation{Key Laboratory of Quark and Lepton Physics (MOE) \& Institute of Particle Physics, Central China Normal University, Wuhan, Hubei 430079, China}

\author{Weiyao Ke}
\email{weiyaoke@ccnu.edu.cn}
\affiliation{Key Laboratory of Quark and Lepton Physics (MOE) \& Institute of Particle Physics, Central China Normal University, Wuhan, Hubei 430079, China}
\affiliation{Southern Center for Nuclear-Science Theory (SCNT), Institute of Modern Physics, Chinese Academy of Sciences, Huizhou, Guangdong 516000, China}

\author{Guang-You Qin}
\email{guangyou.qin@ccnu.edu.cn}
\affiliation{Key Laboratory of Quark and Lepton Physics (MOE) \& Institute of Particle Physics, Central China Normal University, Wuhan, Hubei 430079, China}

\author{Lei Wang}
\email{leiwangqd@sdu.edu.cn}
\affiliation{Key Laboratory of Quark and Lepton Physics (MOE) \& Institute of Particle Physics, Central China Normal University, Wuhan, Hubei 430079, China}

\begin{abstract}
The energy-energy correlator (EEC) and multiplicity are two fundamental observables probing complementary aspects of QCD jets: the former characterizes the angular structure of energy flows in a scale-dependent manner, while the latter is sensitive to the entire history of particle production. In this \emph{Letter}, we uncover a nontrivial correlation between them by studying the EEC as a function of jet internal multiplicity.
We introduce the multiplicity-conditioned EEC jet function (MCJF) and perform a factorization calculation to next-to-leading order accuracy. It is found that, for jet samples selected at a given normalized multiplicity $\nu = N_{\rm ch}/\langle N_{\rm ch} \rangle$, the EEC in the angular region $\Lambda_{\rm QCD}/p_{T,\rm jet}\ll\chi\ll R$ acquires a $\nu$-dependent anomalous dimension. 
Thus the $\nu$-conditioned EEC provides a direct and robust probe to the multiplicity generating function in the perturbative regime. In addition, understanding $\nu$ dependence of the EEC is also crucial for isolating possible multiplicity-dependent bias effects in the EEC measurements in nuclear environment.
\end{abstract}

\maketitle

%%%%%%%%%%%%%%%%%%%%%%%%%%%%%%%%%%%%%%%%%%%%%%%%%%%%%%%%  Introduction 
{\it \noindent Introduction.---} Jet substructures are variable resolution probes of partonic dynamics and enable high-precision tests of perturbative calculations~\cite{Seymour:1997kj,Butterworth:2008iy, Larkoski:2017jix, Marzani:2019hun}. 
The \emph{jet energy correlator (EEC)}, a class of energy-weighted angular correlation functions, stands out as a unique substructure tool~\cite{Basham:1978bw, PhysRevD.24.2383, Chen:2020vvp}. In particular, the EEC admits factorization in both the collinear and back-to-back limits~\cite{Moult:2018jzp, Dixon:2019uzg, Duhr:2022yyp}, offers a scale-dependent scan of QCD radiation, and provides a controlled handle on perturbative and non-perturbative physics and their transition~\cite{Schindler:2023cww, Lee:2024esz, Herrmann:2025fqy}. 
The EEC has thus been studied extensively across $e^+e^-$ annihilation~\cite{Basham:1977iq, Basham:1978zq, PLUTO:1979vfu}, hadronic and nuclear collisions~\cite{Liu:2022wop, Liu:2023aqb, Yang:2023dwc, Barata:2023bhh, Liu:2024kqt, Liu:2024lxy, Xing:2024yrb, Barata:2024wsu, Bossi:2024qho, Chen:2024nfl, Ke:2025ibt, CMS:2024mlf, ATLAS:2023tgo, ALICE:2024dfl, STAR:2025jut}, and deep-inelastic scattering~\cite{Li:2020bub, Li:2021txc, Devereaux:2023vjz, Kang:2023big, Guo:2025qnz}, to understand QCD dynamics in both vacuum and various nuclear medium environments.
{\it The jet internal multiplicity}~\cite{Koba:1972ng, KHOZE1997179, Mueller:1982cq, Dokshitzer:1982xr, Lam:1983vw, Bassetto:1983mvz, Hinz:1985wq, Dokshitzer:1991wu, DOKSHITZER1993295, Hegyi:2000sp, Dokshitzer:2005ri, H1:2020zpd, Vertesi:2020utz, Liu:2022bru, Liu:2023eve, Germano:2024ier, Dokshitzer:2025fky, Duan:2025ngi, Duan:2025lvi, Dremin:1994bj, Malaza:1987hj, GAFFNEY1985109, Duan:2025gpp} is another fundamental observable that drives the understanding of QCD particle production. As a simple count of particles within a jet, it lost the resolution to branchings at different scales, but rather records the cumulative jet activity during its entire evolution history, including both perturbative and non-perturbative stages.

In this \emph{Letter}, we show that beyond the individual development of jet EEC and multiplicity, their correlation is an equally informative probe of QCD partonic dynamics, though it remains unexplored from first principles. 
For this purpose, we study this correlation by introducing the multiplicity-conditioned EEC; it is defined by measuring the EEC within jet samples restricted to a fixed range of particle multiplicity, or the normalized multiplicity $\nu=N_{\rm ch}/\langle N_{\rm ch}\rangle$. 
A straightforward way to understand the origin of such correlation is that conditioning on multiplicity effectively selects jet showers with a given amount of branching activity, thereby modifying the typical amount of energy flow shared among different directions. Importantly, we found that salient features of this correlation can be studied with perturbation theory, offering a theoretically controlled way to diagnose the scale-dependent parton production via energy correlators.

We perform a factorization study of this observable using the multiplicity-conditioned EEC jet
function (MCJF) and calculate it at the next-to-leading (NLO) order. Using the multiplicity generating function formalism, renormalization group equations that govern the co-evolution of EEC and multiplicity are derived. We show that the $\nu$-conditioned EEC in the collinear region retains a power-law form as a function of angle $\chi$, but the exponent acquires a $\nu$ dependence determined by the multiplicity generating function in the perturbative regime.

The establishment of such a correlation is important in two ways. First, it builds a direct, quantitative link between the robust angular power-law exponent and parton production, which is a connection previously only accessible in simulations. Second, it highlights the necessity of accounting for differences in multiplicity distributions when comparing the EEC measurements across different jet samples, particularly in proton-lead or election-ion collisions, where a fluctuating medium background may bias the measured multiplicity distributions relative to those in the vacuum. Using this new tool, we directly study such ``apparent modifications'' of the EEC.

%%%%%%%%%%%%%%%%%%%%%%%%%%%%%%%%%%%%%%%%%%%%%%%%%%%%%%%%%%%%%%%%%%%% Section 1
{\it \noindent The Multiplicity Generating Function.---} 
The theoretical tool to implement multiplicity conditioning of the EEC is the multiplicity generating function, which is a Laplace transformation of a multiplicity distribution $P(N)$: 
\begin{align}
\hskip-.9em Z(s) = \sum_{N=0}^\infty e^{-s N } P(N)\,,\,
P(N) = \int_{s_0-i\pi}^{s_0+i\pi} Z(s) \frac{e^{sN} ds}{2\pi i} \,,
\end{align}
where $s$ is the variable conjugate to $N$. 

For semi-inclusive jets~\cite{Kang:2016mcy,Dai:2016hzf} with the jet energy $\omega_J$ being a fraction $z_J$ of the energy of the initial parton ``$i$'', we denote its semi-inclusive multiplicity distribution and the corresponding generating function as $P_i(z_J,N,\omega_J, R, \mu)$ and $Z_i(z_J, s, \omega_J, R, \mu)$, respectively. $R$ is the jet cone size and $\mu$ is the renormalization scale. 
In the following, we often suppress $\omega_J$, $R$, $\mu$ for brevity unless otherwise needed. 
The NLO calculation and renormalization of $Z_i$ have been given in our earlier work~\cite{Duan:2025gpp}; here we borrow the key results that are relevant for the study of multiplicity conditioned EEC.

If the contamination due to multiplicity produced from partons outside the jet cone is neglected,
the cross section to produce a jet  with a fixed $N$ factorizes,
\begin{align}\label{eq:dsigma_at_fixed_N}
  \frac{\mathrm{d} \sigma_{pp \to J(N)+X}}{\mathrm{d} p_{T,J} \mathrm{d} \phi \mathrm{d} \eta} =& \sum_{i} \int\frac{\mathrm{d} z_J}{z_J}\frac{\mathrm{d} \sigma_{pp\to i}}{\mathrm{d}p_{T,i}\mathrm{d}\phi \mathrm{d} \eta}\left(\frac{p_{T, J}}{z_J}, \mu_H\right)\nonumber\\
  &  \hspace{1.5cm} \times P_i(z_J,N,\omega_J, R, \mu_H)\,,
\end{align}
where $\mathrm{d} \sigma_{pp \to i}$ is the hard partonic cross section~\cite{AVERSA1989105, Jager:2002xm}, and $P_i(z_J,N)$ can be obtained from the inverse Laplace transformation of $Z_i(z_J,s)$. At the jet scale $\mu_J\sim \omega_J R$, the semi-inclusive $Z_i(z_J,s)$ further factorizes into the hard-collinear function $J_{ji}(z_J)$ that encodes out-of-cone radiations, and the exclusive multiplicity generating function $Z_j(s)$ (distinguished by the absence of $z_J$) that describes branching of particles inside the jet,
\begin{align}
\hskip-.5em Z_i(z_J, s, \omega_J, \mu_H) = \sum_{j} J_{ji}(z_J, \mu_H, \mu_J) Z_j(s, \omega_J, \zeta_J)\,.
\end{align}
The time-like DGLAP equations govern the evolution of $J_{ji}(z_J)$ from jet scale $\mu_J$ to hard scale $\mu_H$~\cite{Kang:2016mcy, Kang:2016ehg, Dai:2016hzf, Kang:2017frl}.
The renormalized $Z_i(s,\omega, \zeta)$ follows an angular-ordered evolution to properly treat soft gluon emissions~\cite{Bassetto:1983mvz, Duan:2025gpp}:
\begin{align}
\label{eq:Z-RGE}
  \frac{\partial Z_i (s, \omega, \zeta)}{\partial \ln \zeta} &= S\int_0^1 \mathrm{d} x \frac{\alpha_s(k_{\perp}^2)}{2\pi} \Theta {\left( k_{\perp}^2 - Q_0^2 \right)} \nonumber \\
  \times\sum_{j,k}&\hat{P}_{ji}(x) Z_j(s,x\omega,\zeta) Z_k(s,(1-x)\omega, \zeta) \,.
\end{align}
Here, $\zeta = 1 - \cos \theta$ with $\theta$ being the angle of parton radiation.
$\hat{P}_{ji}(x)$ are leading-order QCD splitting functions~\cite{Altarelli:1977zs, Catani:1996vz}. The symmetry factor $S=(1+\delta_{jk})^{-1}$ accounts for identical-particle effects in the final state.
$Q_0$ is a cutoff to avoid the IR pole of the running coupling. 
$Z_i$ is evolved from its initial data from an IR scale $\zeta_0$ to $\zeta_J$, with $\zeta_J = 1-\cos R$ for consistent matching with hard collinear function at scale $\mu_J$.
The non‑perturbative initial condition $Z(s,\omega,\zeta_0)$ is parameterized as in Ref.~\cite{Duan:2025gpp}.

%%%%%%%%%%%%%%%%%%%%%%%%%%%%%%%%%%%%%%%%%%%%%%%%%%%%%%%%%%%%%%%%%%%%%%%%%%% Section 2
{\it \noindent Jet EEC conditioned on multiplicity.---} The energy-weighted cross section of semi-inclusive jets produced at fixed multiplicity $N$ and  the EEC angle $\chi$ factorizes into:
\begin{align}\label{eq:EECfactorize}
     \frac{\mathrm{d} \Sigma(\chi, N, \omega_J)}{\mathrm{d} \chi \mathrm{d} p_{T,J}}
    =& \sum_i \int \frac{\mathrm{d}z_J}{z_J} \frac{\mathrm{d} \sigma_{pp \to i}}{\mathrm{d} p_{T,i}}\left( \frac{p_{T,J}}{z}, \mu_H\right) \nonumber \\
    &\hskip2em \times \mathcal{G}_i(z_J, \chi, N, \omega_J, R, \mu_H) ,
\end{align}
where $\mathcal{G}_i(z_J,\chi,N)$ is the semi‑inclusive multiplicity-conditioned jet function (SiMCJF) with initiating parton ``$i$''. If one sums over $N$, this definition recovers the standard EEC.
We now investigate the generating function of SiMCJF, i.e., $\mathcal{G}_i(z_J,\chi,s)$ and focus on the collinear limit with a hierarchy $\Lambda_{\text{QCD}}/\omega_J \ll \chi \ll R$. In this limit, radiative corrections fall into three angular regions. 

{\it Region-I~} The first region contains emissions outside of the jet cone $\theta>R$, which lead to the semi-inclusive measurement of the jet. As usual, they factor out as the hard-collinear functions,
\begin{align}\label{eq:factorizeSigma}
\hskip-.5em\mathcal{G}_i(z_J, \chi, s, \omega_J, \mu_H) = J_{ji}(z_J, \mu_H, \mu_J) \mathcal{G}_j(\chi, s, \omega_J, \zeta_J)\,,
\end{align}
with $\mathcal{G}_j(\chi, s, \omega_J, \zeta_J)$ being the generating function of the exclusive MCJF.

{\it Region-II~} We then factor out from $\mathcal{G}_j(\chi, s, \omega_J, \zeta_J)$ emissions in the second angular region: $\chi < \theta < R$. Such emissions take away energy from the measured two-point energy flows, yet still produce particles inside the jet. Therefore, these are the modes responsible for introducing the interested correlation between the EEC and multiplicity. 
Because such radiations are in the perturbative region, we perform an NLO calculation and obtain
\begin{align}
    \mathcal{G}_i(\chi,  s, \omega_J,\zeta_J) &=  \frac{\alpha_s}{\pi} \frac{{(\mu^2 e^{\gamma_E})}^{\epsilon}}{\Gamma(1-\epsilon)} S \int \frac{\mathrm{d} k_{\perp}}{k_{\perp}^{1+2\epsilon}} \int \mathrm{d} x \hat{P}_{ji}(x, \epsilon)  \nonumber \\
      & \hskip-2.2cm \times\Theta_{\text{alg}}\big\{x(1-x)\omega_J^2\delta(\chi-\theta_{jk}) Z_j(s,x\omega_J) Z_k(s,(1-x)\omega_J)\nonumber\\
    &\hskip0cm + \mathcal{G}_j(\chi, s, x\omega_J) Z_k(s, (1-x)\omega_J)\nonumber \\
    &\hskip0cm + Z_j(s,x\omega_J)\mathcal{G}_k(\chi, s, (1-x)\omega_J)~\big\}\,. 
\end{align}
$\hat{P}_{ji}(x, \epsilon)$ is the splitting function in $d=4-2\epsilon$ dimensions~\cite{Catani:1996vz}, $\Theta_{\text{alg}}$ is jet algorithm constraint (anti-$k_T$ in this study~\cite{Cacciari:2008gp}). The three terms in curly brackets represent the non-contact and contact contributions to the EEC, respectively. The additional factors of multiplicity generating function $Z_i(s)$ keeps track of the particle production within the jet. 

Using the renormalization procedure in Ref.~\cite{Duan:2025gpp}, the renormalized $\mathcal{G}_i$ at the angular scale $\zeta_{\chi}=1-\cos\chi$ is 
\begin{align}\label{eq:ICEEC}
    &\mathcal{G}_i(\chi, s, \omega_J, \zeta_\chi) = \frac{\omega_J^2}{\chi} S \int \mathrm{d} x  \frac{\alpha_s(k_{\perp}^2) }{\pi} \Theta(k_{\perp}^2 - Q_0^2) \nonumber \\
     &\hspace{1.5cm} \times \sum_{j,k} x(1-x) \hat{P}_{ji}(x) Z_j(x) Z_k( 1-x) \,.
\end{align}
Here, the shorthand notations $\mathcal{G}_j(x) \equiv \mathcal{G}_j(\chi, s, x\omega_J, \zeta)$ and $Z_j(x)\equiv Z_j(s,x\omega_J,\zeta)$ are used.
From such an initial condition, $\mathcal{G}_i(\chi, s, \omega_J, \zeta_J)$ can be obtained by solving the following angular‑ordered RGE:
\begin{align}
    &\frac{\partial \mathcal{G}_i(\chi, s, \omega_J, \zeta)}{\partial \ln \zeta} = S\int \mathrm{d} x \frac{\alpha_s(k_{\perp}^2) }{2\pi} \Theta(k_{\perp}^2 - Q_0^2) \nonumber \\
    & \times \sum_{j,k} \hat{P}_{ji}(x) {\big\{~ \mathcal{G}_j(x)  Z_k(1-x) + \mathcal{G}_k(1-x) Z_j(x) ~\big\}}\,.
\end{align}
It should be noted that because the measurement of energy flux on one parton does not affect the particle production on the unmeasured parton, the multiplicity generating function $Z_i$ in the above equation evolves independently according to Eq.~(\ref{eq:Z-RGE}). As a result, the RGEs for $\mathcal{G}_i(\chi, s, \omega_J, \zeta)$, unlike the ones for $Z_i(s)$, remain linear differential-integral equations. This means that the EEC still acquires the power-law form even with multiplicity conditioning. In fact, setting $s=0$ removes the multiplicity conditioning (since $Z(s=0) = 1$) and the above equation recovers the equation that sets the anomalous dimension for the standard EEC. Any deviations of the multiplicity-conditioned EEC from the standard power-law exponents can be attributed to the appearance of the multiplicity generating function at scales from $\zeta_\chi$ to $\zeta_R$. This is the reason why the multiplicity-conditioned EEC can be used to diagnose scale-dependent particle production, or more specifically, $Z(s,\omega, \zeta)$ itself.

{\it Region-III~} Finally, in the third region, emissions have angles much smaller than the EEC angle $\theta\ll\chi$; they cannot change the energy flow at leading power and only contribute to multiplicity production. Therefore, the inclusion of such modes is achieved by solving $Z_i$ at scale $\zeta_\chi$ from its NP initial condition at $\zeta_0$ via Eq.~(\ref{eq:Z-RGE}).

\begin{figure}[htbp]
  \begin{center}
    \includegraphics[scale=0.44]{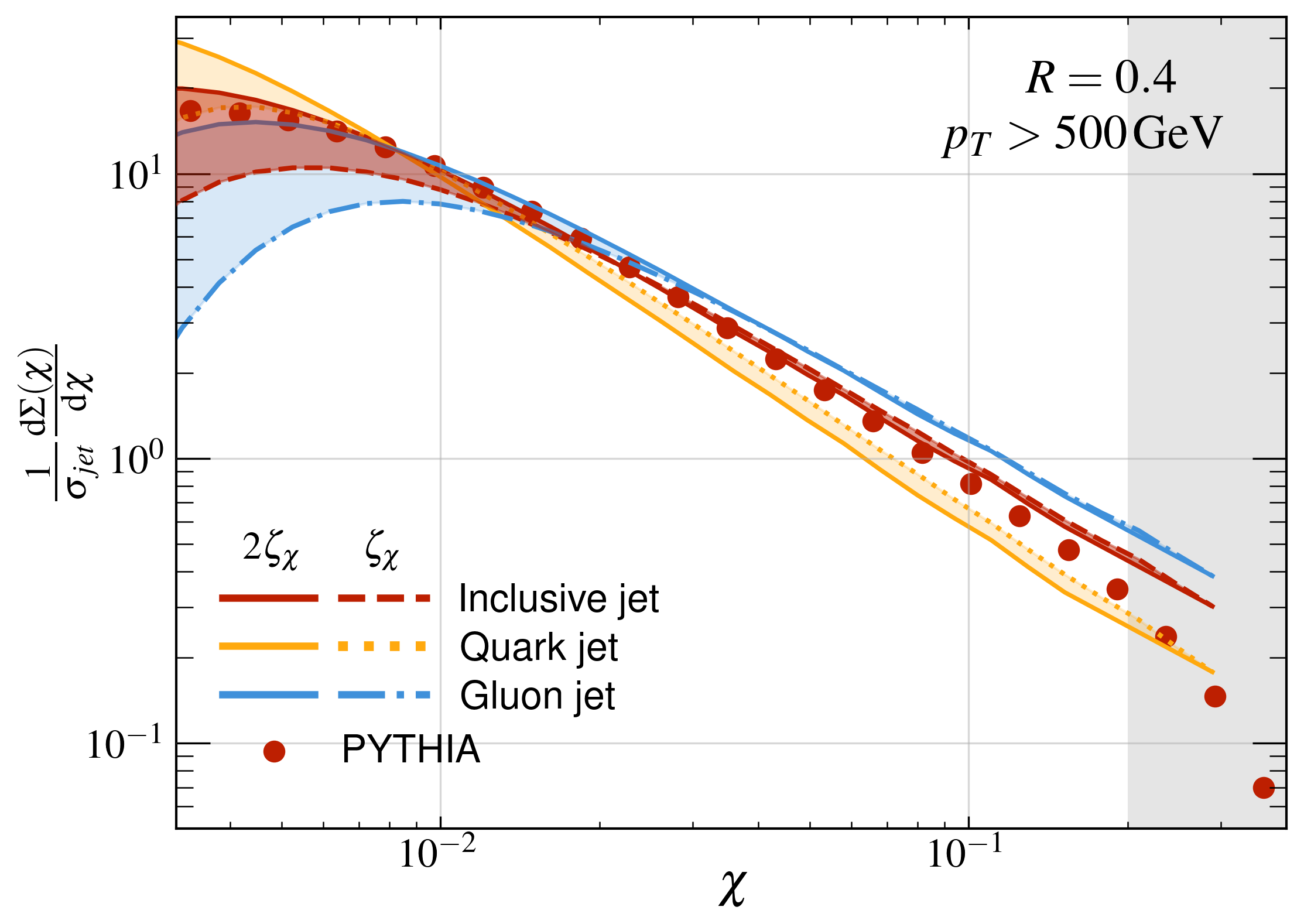}
    \caption{Standard EEC for jets with $p_T > 500$ GeV and $R = 0.4$. LO+LL predictions (bands) are compared to \textsc{Pythia8} simulations (markers). The bands represent the perturbative uncertainty from varying the matching scale between $\zeta_{\chi}$ and $2\zeta_{\chi}$. Results are shown for inclusive (red), quark (orange), and gluon (blue) jets.}
    \label{fig:eec_multi_vs_no}
 \end{center}
\end{figure}

%%%%%%%%%%%%%%%%%%%%%%%%%%%%%%%%%%%%%%%%%%%%%%%%%%%%%%%%%%%%%%%%%%%%%%%%%%%%%%%%%%%%%%%%%% Results
{\it \noindent Consistency check for standard EEC.---} As a validation of our framework, we first verify that summing the multiplicity-conditioned EEC over all multiplicities reproduces the standard inclusive EEC. This consistency check is satisfied by construction in our formalism, and we explicitly confirm it numerically. In Fig.~\ref{fig:eec_multi_vs_no}, we present our LO+LL predictions for the standard EEC, shown as uncertainty bands obtained by varying the matching scale between $\zeta_\chi$ and $2\zeta_\chi$, and compare them with \textsc{Pythia8}~\cite{Bierlich:2022pfr} simulations for jets with $p_T > 500$ GeV and cone radius $R = 0.4$. The inclusive EEC from our calculation agrees well with \textsc{Pythia8} across a wide angular range, except for $\chi \lesssim 10^{-2}$, where non-perturbative effects become important. The gluon-jet EEC exhibits a harder angular spectrum, consistent with its larger anomalous dimension~\cite{Dixon:2019uzg}. This agreement provides a first sanity check before delving into the multiplicity-conditioned EEC.

\begin{figure}[htbp]
\centering
\includegraphics[scale=0.42]{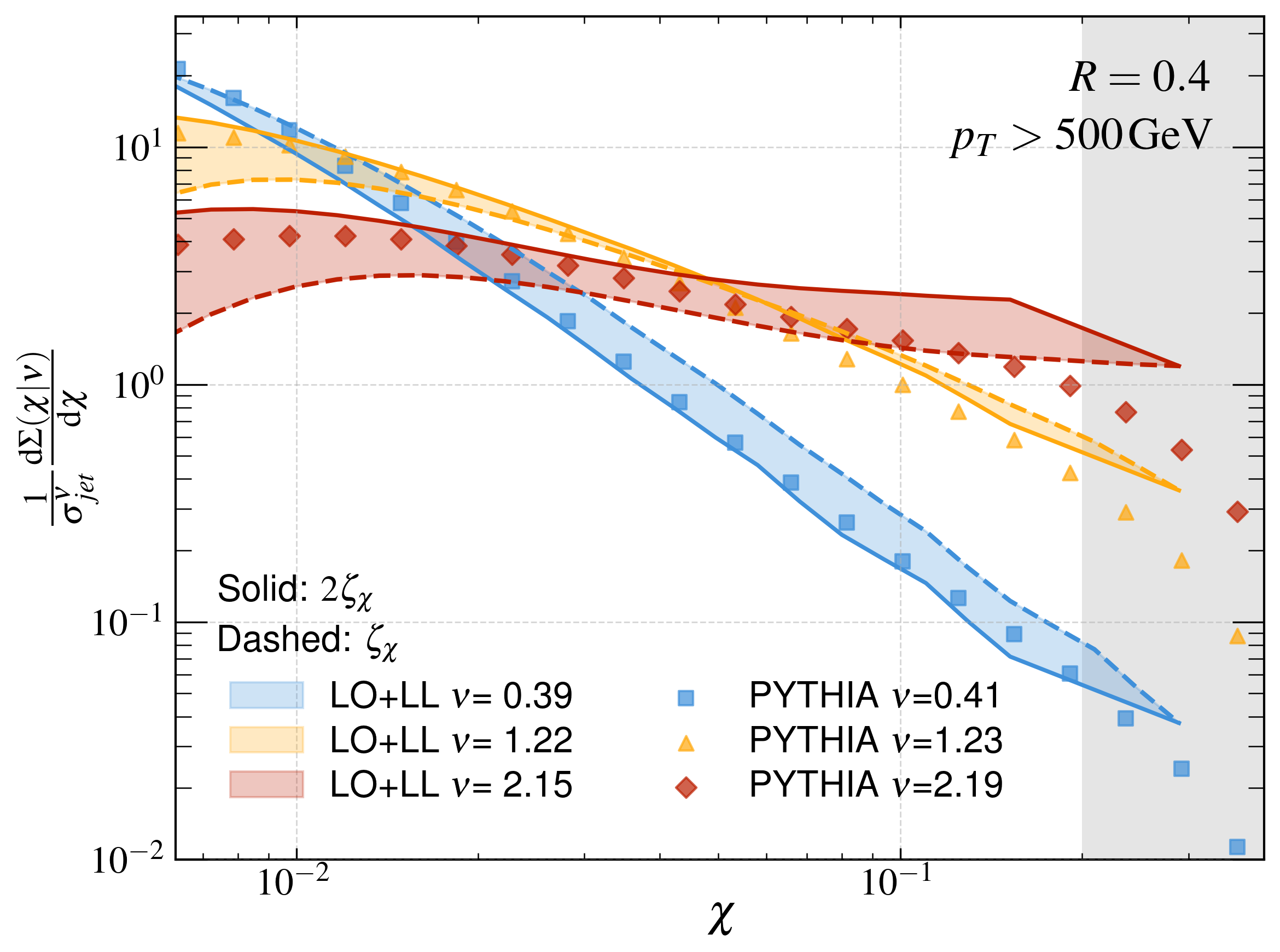}
\caption{Multiplicity-conditioned EEC for jets with $p_T > 500$ GeV and $R = 0.4$, compared with \textsc{Pythia8} simulations for different multiplicity classes $\nu = N_{\text{ch}}/\langle N_{\text{ch}} \rangle$. $\sigma_{\text{jet}}^{\nu}$ is the cross section for jet with normalized multiplicity $\nu$. Bands show LO+LL predictions with the matching scale varied between $\zeta_{\chi}$ and $2\zeta_{\chi}$, markers denote \textsc{Pythia8} results.}
\label{fig:eec_multi}
\end{figure}

{\it \noindent The $\nu$-conditioned EEC.---}
The EEC conditioned on a specific multiplicity $N$ is obtained as
\begin{align}
\frac{\mathrm{d} \Sigma(\chi,\omega_J|N_{\rm ch})} {\mathrm{d} \chi} = \frac{1}{P(N_{\rm ch},\omega_J)}\frac{\mathrm{d} \Sigma( \chi,N_{\rm ch},\omega_J)}{\mathrm{d} \chi} \,,
\end{align}
where $P(N_{\rm ch},\omega_J)=d\sigma_N/d\sigma_{\text{jet}}$ can be obtained from Eq.~(\ref{eq:dsigma_at_fixed_N}). We present the result in different range of the normalized multiplicity classes $\nu = N_{\text{ch}}/\langle N_{\text{ch}} \rangle$. 

As shown in Fig.~\ref{fig:eec_multi}, the LO+LL predictions reproduce the \textsc{Pythia8} results well across the shown $\nu$ range and reveal a systematic flattening of EEC as $\nu$ increases. This flattening can be understood as follows. Emissions in the angular range between $\chi$ and $R$ suppress the EEC by draining energy from the prongs that contribute at angle $\chi$. This suppression is stronger for higher‑multiplicity jets, as the samples are biased toward jets with increased radiation activity, leaving less energy available to be shared among energy flows separated by a small $\chi$. Conversely, the increased number of splittings also produces more particle pairs separated at larger angles, enhancing the EEC there. 

Remarkably, the \textsc{Pythia8} simulation of the EEC within different $\nu$ ranges indeed retains power‑law behavior in the region $\Lambda_{\rm QCD}/\omega_J \ll \chi \ll R$, i.e., $\Sigma(\chi|\nu) \sim \chi^{\gamma(\nu)}$. The anomalous dimension, or effective exponent $\gamma(\nu)$, acquires a $\nu$-dependence. This is expected from the theoretical analysis, as the evolution equation for $\mathcal{G}_i$ is linear with $Z_i$ as independent inputs. 

\begin{figure}[!t]
  \begin{center}
    \includegraphics[scale=0.44]{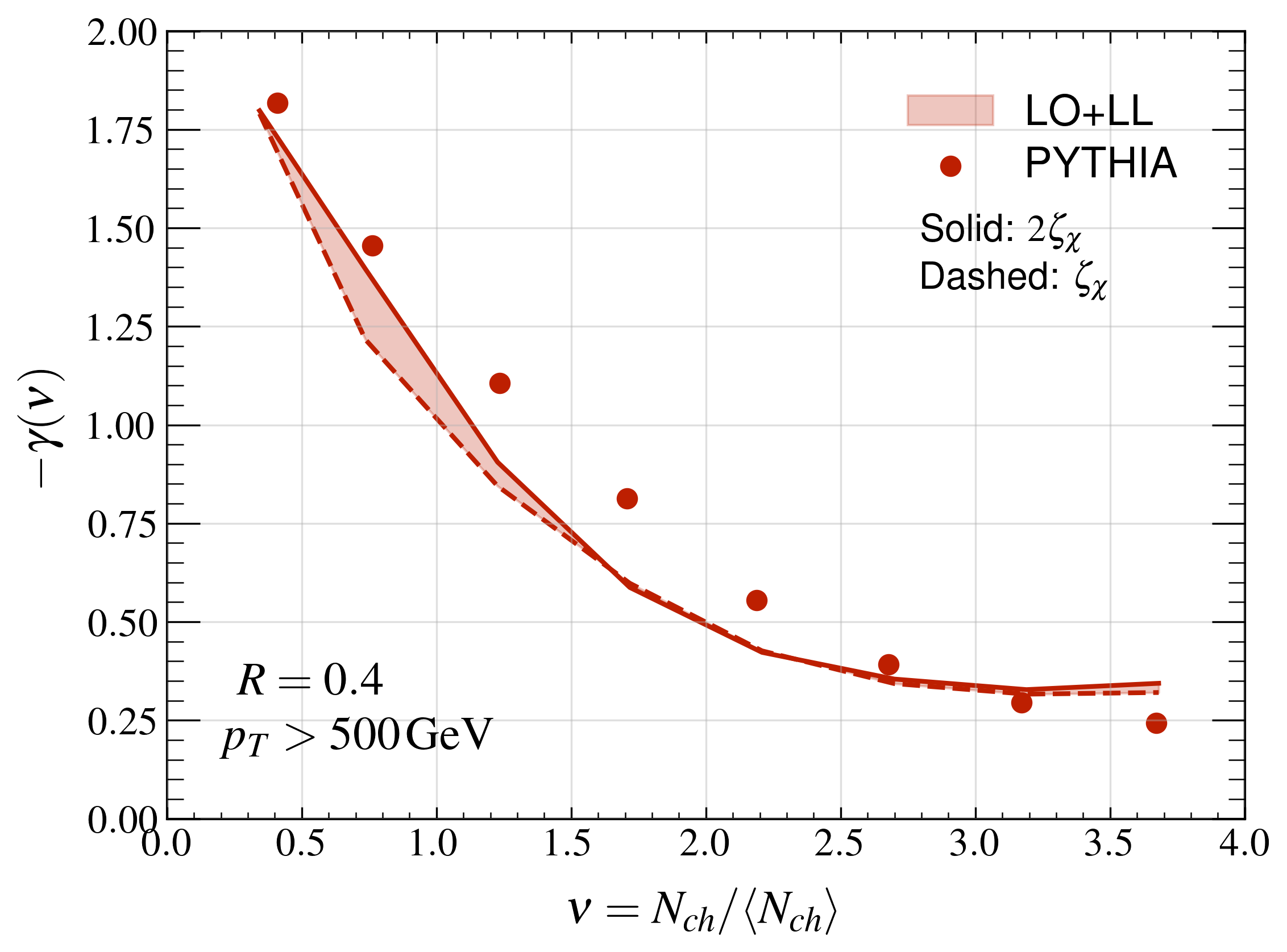}
    \caption{Exponent of the multiplicity-conditioned EEC in the asymptotic region as a function of $\nu = N_{\text{ch}} / \langle N_{\text{ch}} \rangle$ for jets with $p_T > 500$ GeV and $R = 0.4$. \textsc{Pythia8} results are shown as symbols; LO+LL predictions as bands.}
    \label{fig:eec_slope}
 \end{center}
\end{figure}

In Fig.~\ref{fig:eec_slope}, we extract the exponent $\gamma(\nu)$ and show it as a function of $\nu$. Both the LO+LL calculation and \textsc{Pythia8} exhibit a clear monotonic decrease of the slope with increasing $\nu$. While the LO+LL prediction does not fully agree with \textsc{Pythia8} simulations, it still captures the essential trend.

Given that exponents are robust features extractable from experiments, $\gamma(\nu)$ can be used to diagnose particle production via the EEC. Conditioning the EEC on $\nu$ selects jets with a specific amount of branching activity, and the resulting exponent reveals, in a scale‑sensitive manner, how the parton shower activity encoded in the multiplicity influences the angular distribution of energy. This makes the $\nu$-conditioned EEC a novel probe of the perturbative dynamics underlying multiple particle production, complementary to traditional studies of multiplicity distributions or inclusive jet substructure.

\begin{figure}[htbp]
  \begin{center}
    \includegraphics[scale=0.6]{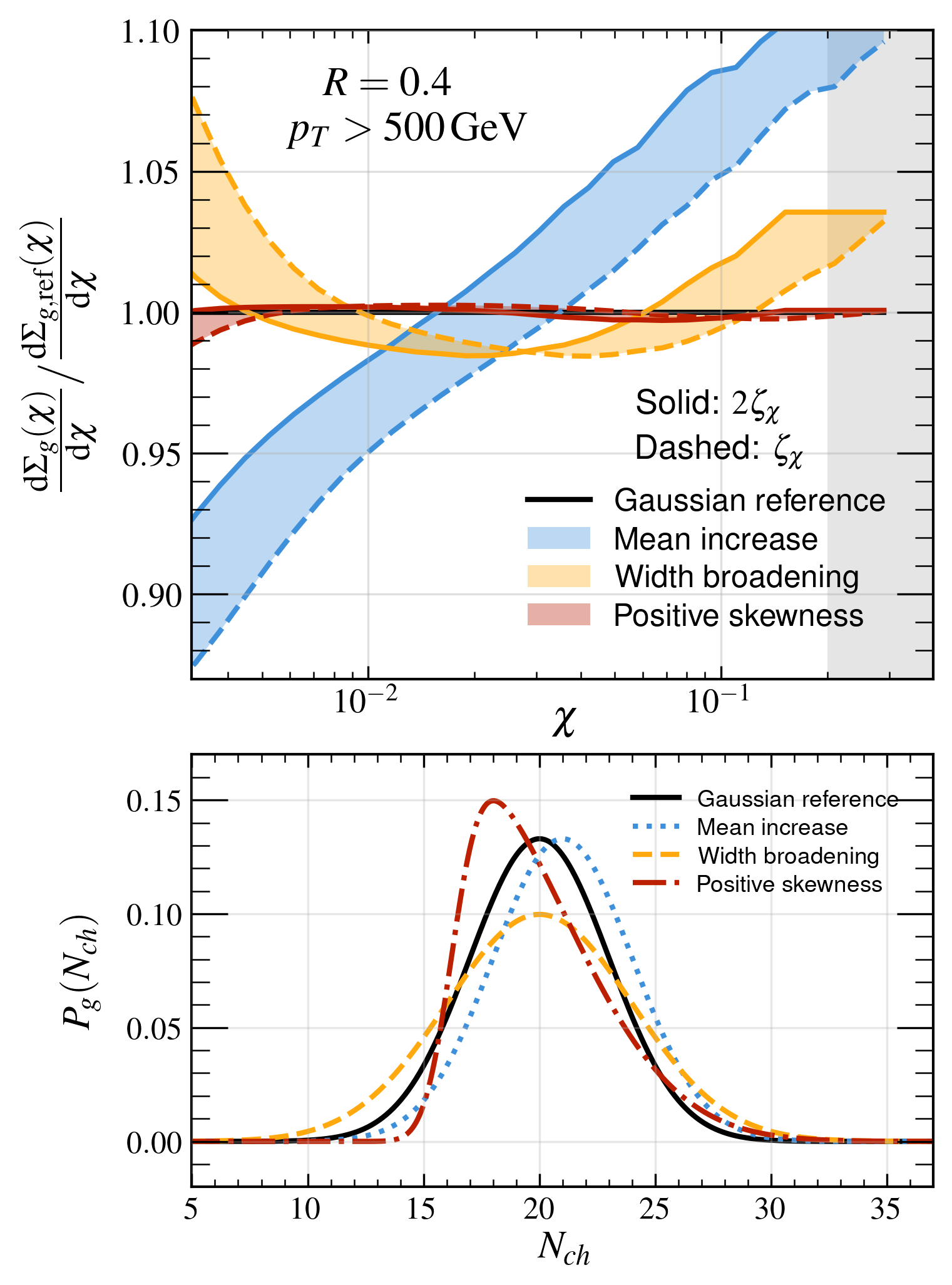}
    \caption{Ratios of the standard gluon‑jet EEC obtained with different multiplicity distributions $P_g(N_{\text{ch}})$, modeled using a skew‑normal ansatz. Variations in the mean (blue dotted), width (yellow dashed), and skewness (red dash‑dotted) are shown relative to a Gaussian reference (black solid).}
    \label{fig:eec_ratio}
 \end{center}
  \vspace{-5.ex}
\end{figure}    

{\it \noindent Impact of multiplicity bias on EEC.---} 
Because the EEC exhibits a systematic dependence on $\nu$, collisions involving a highly fluctuating background, like in reactions involving nuclei, can bias the jet measurement reuslting in a modifyed multiplicity distribution of the measured jet samples and thereby induce apparent modifications to the EEC. Using the tools we developed, we compute the EEC averaged over jet samples with different multiplicity distributions and demonstrate this effect.

Using gluon jets as an example, for a jet samples with given multiplicity distribution, we can compute the ensemble average EEC as
\begin{align}
\frac{\mathrm{d} \Sigma_g(\chi, \omega_J)}{\mathrm{d} \chi} = \sum_{N_{\text{ch}}} P_g(N_{\text{ch}}, \omega_J) \frac{\mathrm{d} \Sigma_g(\chi, \omega_J|N_{\text{ch}})}{\mathrm{d} \chi}.
\end{align}
For illustration purpose, we uses a skew‑normal ansatz for $P_g(N_{\text{ch}})$, which is controlled by the mean , the width (standard deviation), and the skewness.
Four distributions are tested as shown in the lower panel of Fig.~\ref{fig:eec_ratio}.
The first case is a Gaussian distribution (zero skewness, black-solid line) with the mean $\overline{N}_{\text{ch}}=20$ and width $\sigma=3$ as the reference.
If one increase the mean (blue dotted line), the resulting ratio of its EEC relative to the reference case is apparently modified (blue band in the upper panel of Fig.~\ref{fig:eec_ratio}). 
It is tilted towards large angles, which is signature of the flattening associated with higher‑multiplicity jets. Widening the distribution from $\sigma =3$ to $\sigma=4$ while keeping the mean fixed (yellow lines and bands) enhances both small- and large‑angle regions relative to the reference, reflecting the increased weight of both low‑ and high‑multiplicity jets.
Finally, introducing a moderate positive skewness produces only mild distortions (red lines and bands).

%%%%%%%%%%%%%%%%%%%%%%%%%%%%%%%%%%%%%%%%%%%%%%%%%%%%%%%%%%%%%%%%%%%%%%%%%%%%%%%%%%%%%%%%%% Conclusion 
{\it \noindent Conclusion.---} 
We investigate the correlation between the jet energy-energy correlator (EEC) and the internal multiplicity of jets. We introduced the multiplicity conditioned jet function (MCJF, and its associated generating function). The semi-inclusive MCJF factorizes in the limit $\Lambda_{\rm QCD}/\omega_J \ll \chi \ll R$, and we compute it to next-to-leading order accuracy.
It is found that non-trivial correlations between EEC and multiplicity are generated by radiations with angles between $\chi$ and $R$, and a set of joint evolution equations is developed to resum the leading logarithmic corrections to this correlation.  The central result is the emergence of a multiplicity-dependent anomalous dimension of EEC in the collinear region. Calculations show good agreement with \textsc{Pythia8} simulations as functions of normalized multiplicity $\nu=N_{\rm ch}/\langle N_{\rm ch}\rangle$, demonstrating a good perturbative control.
This establishes a robust, analytic connection between angular pattern of energy flow and particle production. The $\nu$-conditioned EEC therefore offers a novel probe of the scale-dependent dynamics underlying multiple particle production.

Beyond its theoretical interest, this correlation provides important implications for experimental data interpretation. Because the EEC exhibits a sensitivity to multiplicity conditioning, apparent modifications in ratios such as $pA/pp$ may arise from multiplicity biases induced by the collision environment. These findings underscore the necessity of carefully controlling or accounting for multiplicity selection effects when interpreting EEC measurements across different collision systems.

By bridging two fundamental jet observables, this work opens a new direction for QCD parton shower studies, providing a theoretically controlled handle on the interplay between IRC safe jet substructure and particle production. Future work includes extending the study to higher accuracy, incorporating non-perturbative corrections, and applying it to more complicated collision environments, such as $pA$ and $eA$.

\textbf{\textit{Acknowledgment.}}
The authors would like to thank Lin Chen, Hai-Tao Li, and Wenbin Zhao for helpful discussions. This work is supported in part by National Natural Science Foundation of China (NSFC) under Grant Nos. 12225503, 1234710148, and 12575140, and in part by China Postdoctoral Science Foundation under Grant No. 2023M742098. Some of the calculations were performed in the Nuclear Science Computing Center at Central China Normal University ($\mathrm{NSC}^3$), Wuhan, Hubei, China. The authors acknowledge the use of AI-based writing tools to assist with language refinement and editing.

\bibliographystyle{apsrev4-1}
\bibliography{refs.bib}

\end{document}